\documentclass[12pt]{article}

\usepackage{amsfonts}

\begin{document}

%%%%%%%%%%
\title{Gauge symmetry in Fokker-Planck dynamics }

\author{M. de Montigny$^{a,b}$, F.C. Khanna$^{b,c}$ and A.E. Santana$^{b,d}$\\
$^{a}$Facult\'{e} Saint-Jean, University ofAlberta \\
8406-91 Street, Edmonton, Alberta, CanadaT6C 4G9 \\
$^{b}$Theoretical PhysicsInstitute, University of Alberta\\
Edmonton, Alberta, Canada T6G 2J1 \\
$^c$TRIUMF, 4004, Wesbrook Mall \\
Vancouver, British Columbia, Canada V6T 2A3\\
$^{d}$Instituto de F\'{i}sica, Universidade Federal da Bahia\\
Campus de Ondina, Salvador, Bahia, Brazil 40210-340}

\maketitle
 
\begin{abstract}
Using a Galilean metric approach, based in an embedding of the Euclidean
space into a (4+1)-Minkowski space, we analyze a gauge invariant Lagrangian
associated with a Riemannian manifold ${\cal R}$, with metric $g$. With a
specific choice of the gauge condition, the Euler-Lagrange equations are
written covariantly in ${\cal R}$, and then the Fokker-Planck equation is
derived, such that the drift and the diffusion terms are obtained from $g.$
The analysis is carried out for both, abelian and non abelian symmetries,
and an example with the $su(2)$ symmetry is presented.
\end{abstract}

\footnote{%
E-mail: montigny@phys.ualberta.ca, \ khanna@phys.ualberta.ca, \
santana@ufba.br\ (corresponding Author)}

\newpage

\section{Introduction}

In this paper we show that the Fokker-Planck equation can be derived via a
gauge invariant theory. The basic ingredient in the derivation is Galilean
covariance, which has been recently developed in different perspectives,
providing a metric, and thus a tensor, structure for non relativistic theory
based in a 4+1 Minkowski space\cite
{takahashi1,takahashi2,omote,kunzle,taka3,marc111,marc112,fluidspaper}. As a
consequence, a geometric unification of the non relativistic and
relativistic physics is accomplished\cite{kunzle,marc111}. One interesting
result is that the possibility to use ideas and concepts of particles
physics in transport theory, such as topological terms, symmetry breaking,
gauge symmetries, and so on\cite{takahashi1,takahashi2,jackiw}, can be
investigated in a systematic and covariant way paralleling the relativistic
physics\cite{kunzle,luc}. In this context it would be of interest to analyze
typical stochastic processes such as those described by the Fokker-Planck
dynamics.

The Fokker-Planck equation is often derived in the analysis of Markov
processes. From a physical standpoint, it can be introduced either as the
distribution of probability version of the Langevin equation, describing a
classical particle under the influence of dissipative and stochastic forces 
\cite{risken,tania,gard1}, or as an approximation of the Boltzmann equation 
\cite{liboff}. In this latter case, the collision term is approximated to
consider the transition rate, say $W(p_{1},k)$, where $p_{1}=p+k$, terms up
to the second order in $k$, resulting then in the drift and the diffusion
terms of the Fokker-Planck equation\cite{brian1}. Here we proceed in a
different way, by analyzing (first) a $U(1)$ gauge invariant Lagrangians, in
the (4+1)-dimension Minkowski space (to be referred to as ${\cal G}$). Using
a suitable gauge condition and a proper definition of each component for the
gauge field, the Euler-Lagrange equations result in the Fokker-Planck
equation. The definition of the gauge field is based on the existence of a
Riemannian manifold, say ${\cal {R}({G})}$, with metric $g$, in which ${\cal %
G}$ is taken as a local flat space. Taking the 5-dimensional equations
covariantly written in ${\cal {R}({G})}$, the gauge field is defined with
the use of the metric tensor, which gives rise to the drift and diffusion
terms of the Fokker-Planck equation. The analysis of the connection, defined
by $g,$ establishes whether the diffusion tensor is a constant or not by a
proper coordinate transformation. These results, in addition to improving
the study of symmetries of the Fokker-Planck systems \cite{nos}, opens the
possibility to include in the description of stochastic processes
non-abelian gauge symmetry. This aspect is developed here by using, in
particular, the $SU(2)$ gauge symmetry, following the methods of field
theory, rather than the generalization of symplectic structures and
Liouville equation\cite{heinz}.

The presentation is organized as  it follows. In Section II, to make the
presentation self contained and to fix the notation, a brief outline on the
Galilei covariance is presented. The Fokker-Planck equation is derived from
an-abelian gauge invariant Lagrangian in Section III, and in Section IV the
non-abelian situation is addressed. Final concluding and remarks are
presented in Section V.

\section{Outline on the Galilei Covariance  }

Let us begin with a brief outline of the Galilean covariant methods (for
more details see for instance Ref. \cite{taka3,marc111}). Let ${\cal G}$ be
a five dimensional metric space, with an arbitrary vector denoted by $%
x=(x^{1},x^{2},x^{3},x^{4},x^{5})$ = $({\bf x},x^{4},x^{5})$. The inner
product in ${\cal G}$ is then defined by 
\begin{equation}
(x|y)={\eta }_{\mu \nu }x^{\mu }y^{\nu }=\sum_{i=1}^{n=3}{x^{i}y^{i}}%
-x^{4}y^{5}-x^{5}y^{4},  \label{pescal}
\end{equation}
with $x,y\ \in {\cal G}$ and $\eta _{\mu \nu }$ being given by 
\begin{equation}
\eta =\delta _{ij}dx^{j}{\otimes }dx^{i}-dx^{4}{\otimes }dx^{5}-dx^{5}{%
\otimes }dx^{4}.  \label{galmetri}
\end{equation}

The set of linear transformations in ${\cal G}$ of \ the type $\ {\bar{x}}%
^{\mu }=G_{\ \nu }^{\mu }x^{\nu }+a^{\mu }\ $ (that leaves $(dx|dy)$
invariant), such that $|G|=1$, with $G_{\ \nu }^{\mu }=\delta _{\ \nu }^{\mu
}+\epsilon _{\ \nu }^{\mu }$, admits 15 generators of transformations, and
11 of them provide the Lie Galilei algebra with the usual central extension,
describing the mass of a particle, being a generator of the group.

Consider now the embedding of the Euclidean space ${\cal E}$ in ${\cal G}$,
given by 
\begin{equation}
{\bf A}\ \mapsto \ A=({\bf A},A_{4},\frac{{{\bf A}^{2}}}{{2A}_{4}}),
\label{emb1}
\end{equation}
where ${\bf A=(}A^{1},A^{2},A^{3})\,\in {\cal E}\,,\,A\,\in {\cal G}$. It
follows that $A$ is a null-like vector, since 
\begin{eqnarray*}
(A|A) &=&{\eta }_{\mu \nu }A^{\mu }A^{\nu }\  \\
\ &=&\sum_{i=1}^{3}{A^{i}A^{i}}-2A^{4}A^{5}=0.
\end{eqnarray*}
In other words, according to equation (\ref{emb1}), each vector in ${\cal E}$
is in homomorphic correspondence with null-like vectors in ${\cal G}$. As an
example, consider $x=({\bf x},kt,{\bf x}^{2}/2kt)$, where $k$ is a constant
with units of velocity (we consider $k=1$). Under the subgroup of linear
transformation in ${\cal G}$, given by the generators 
\[
K_{i}=e^{-v^{i}B_{i}},\ \ R_{ij}=e^{\epsilon _{ijk}L_{k}},\ \
T_{i}=e^{a^{i}P_{i}},T_{4}=e^{bH}, 
\]
where $a^{5}=0,\ \ H=P^{4}$, the vector $x=({\bf x},t,{\bf x}^{2}/2t)$
transforms as a Galilean vector; that is 
\begin{eqnarray}
{\bar{x}}^{i} &=&R_{\ j}^{i}x^{j}-v^{i}x^{4}+{\bf a,}  \label{gal1} \\
{\bar{x}}^{4} &=&x^{4}+b,  \label{gal2} \\
{\bar{x}}^{5} &=&x^{5}-v^{i}(R_{\ j}^{i}x^{j})+{\frac{1}{2}}{\bf v}^{2}x^{4}.
\label{gal3}
\end{eqnarray}

There are at least two other types of embeddings, which will be useful here. 
\begin{equation}
{\bf A}\ \mapsto \ A=({\bf A},A_{4},0)  \label{emb2}
\end{equation}
and 
\begin{equation}
{\bf A}\ \mapsto \ A=({\bf A},\frac{1}{\sqrt{2}}A_{4},\frac{1}{\sqrt{2}}%
A_{4}).  \label{emb3}
\end{equation}

We take ${\cal G}$ to be a frame of the following form, 
\begin{equation}
F=dJ+\lambda J\wedge J,  \label{gau121}
\end{equation}
satisfying the Bianch identities $dF+\lambda \lbrack J,F]=0$ and the
equation regarding the sources which are considered to be nonexistent, that
is $d\ast F+\lambda \lbrack J,\ast F]=0$. 

\section{Abelian Gauge symmetry}

We  consider    $\lambda =0$ in Eq. (\ref{gau121}) and write the following $%
U(1)$-gauge invariant Lagrangian, in terms of the components of $F$ (say $%
F^{\mu \nu }$),

\begin{equation}
{\cal {L}}=-\frac{1}{4}F^{\mu \nu }F_{\mu \nu },  \label{lagr1}
\end{equation}
where $F^{\mu \nu }$ is written in terms of the abelian gauge fields as 
\begin{equation}
F_{\mu \nu }=\partial _{\mu }J_{\nu }-\partial _{\nu }J_{\mu },
\label{gau12}
\end{equation}
where $J$ remains to be specified. Following the usual procedure, we can
also write down the Lagrangian as 
\begin{equation}
{\cal {L}}=\frac{1}{2}(\partial _{\mu }J_{\nu }\partial ^{\nu }J^{\mu
}-\partial _{\mu }J_{\nu }\partial ^{\mu }J^{\nu }),  \label{las}
\end{equation}
resulting in the Euler-Lagrange equations 
\begin{equation}
\partial ^{\mu }\partial _{\mu }J^{\nu }-\partial ^{\nu }\partial _{\mu
}J^{\mu }=0.  \label{eul121}
\end{equation}

The Lagrangian ${\cal {L}}$ is invariant under the gauge transformation $%
J^{\mu }\rightarrow \bar{J}^{\mu }=J^{\mu }+\partial ^{\mu }{h(x)}$, and in
the ordinary procedure, we take $\partial _{\mu }J^{\mu }=0$ as the
(Lorentz) gauge condition in order to derive, from Eq. (\ref{eul121}), the
wave equation for the electromagnetic field, that is $\partial ^{\mu
}\partial _{\mu }J^{\nu }=0.$ Here, we are not interested in interpreting $%
J^{\nu }$ as a vector potential, so that we have the freedom to explore a
different gauge condition. We take then the gauge condition to be $\partial
^{\mu }\partial _{\mu }J^{\nu }=0$, such that $h(x)$ fulfills the constraint
equation $\partial ^{\mu }\partial _{\mu }h(x)=\beta $, where $\beta $ is a
arbitrary constant. As a result $\partial _{\mu }J^{\mu }=\alpha $, where $%
\alpha $ is another arbitrary constant, which can be assumed to be zero. The
Euler-Lagrange equations can then be written as 
\begin{equation}
\partial _{\mu }J^{\mu }=0.  \label{f1}
\end{equation}

In order to specify the 5-dimensional vector field theory, we assume the
existence of a Riemannian manifold, ${\cal {R}({G})}$, with metric $g^{\mu
\nu }(x)$, such that at each point of ${\cal {R}({G})}$ there is a flat
space ${\cal {G}}$. The covariant form of Eq. (\ref{f1}) is 
\begin{equation}
\partial _{\mu }(g^{1/2}J^{\mu })=0,  \label{f111}
\end{equation}
such that $J^{\mu }$ is considered as a covariant current density in ${\cal {%
R}({G})}$. We can construct $J$ as an explicit derivative of a tensor of the
theory; and the natural candidate for such a proposal is $g^{\mu \nu }(x)$.
In this way the physical content of $J$ as a current can be emphasized.
Using a general expression for a (covariant-like) derivative, say $\partial
_{\nu }+f_{\nu }(x)$, we define $J^{{\mu }}=g^{-1/2}S^{\mu }$, where $S^{\mu
}=(f_{\nu }(x)+\partial _{\nu })g^{\mu \nu }(x)$, with $f_{\nu }$ being a
5-vector given by $f^{\nu }(x)=(f^{i}({\bf x},t),f^{4}({\bf x},t),0)$ (we
have taken $f^{5}({\bf x},t)=0$ for sake of convenience). The 5-vector $S$
plays the role of the current density in the Minkowski space, and $J$ the
covariant gravitational counterpart. Nevertheless, it is worth noting that
here we are working in the 5-dimensional space ${\cal G}$ without use of the
equivalence principle of the general relativity. The physical meaning of the
Riemannian manifold ${\cal {R}({G})}$ will be discussed in the following.

Using Eq. (\ref{f111}) we find 
\begin{eqnarray}
\partial _{\mu }S^{\mu } &=&\partial _{\mu }f^{\mu }+\partial _{\mu
}\partial _{\nu }g^{\mu \nu }  \nonumber \\
&=&\partial _{4}f^{4}+\partial _{i}f^{i}+\partial _{\mu }\partial _{\nu
}g^{\mu \nu }=0.  \label{f112}
\end{eqnarray}
This equation can be converted into a Fokker-Planck equation if we define: 
\begin{eqnarray*}
f^{i}({\bf x},t) &=&D^{i}({\bf x},t)P({\bf x},t), \\
f^{4}({\bf x},t) &=&P({\bf x},t),
\end{eqnarray*}
and the metric tensor as 
\begin{equation}
g=P(x)D_{ij}(x)dx^{j}{\otimes }dx^{i}-dx^{4}{\otimes }dx^{5}-dx^{5}{\otimes }%
dx^{4}.  \label{metric11}
\end{equation}
where $P(x)=P({\bf x},t)$ is a scalar function and $D^{ij}(x)$ are the
components of a Riemannian metric associated with the Euclidean space. The
components of $S$ are given by 
\begin{eqnarray*}
S^{i} &=&D^{i}(x)P(x)+\partial _{j}P(x)D^{ij}(x), \\
S^{4} &=&P(x), \\
S^{5} &=&0.
\end{eqnarray*}
Using Eq. (\ref{f112}), with the embedding $x=({\bf x},t,{\bf x}^{2}/2t)$,
we obtain 
\begin{equation}
\partial _{t}P({\bf x},t)=\frac{\partial }{\partial x^{i}}\left[ -D^{i}({\bf %
x},t)P({\bf x},t)+\frac{\partial }{\partial x^{j}}D^{ij}({\bf x},t)P({\bf x}%
,t)\right] .  \label{f21}
\end{equation}
This is the Fokker-Planck equation with $D^{i}({\bf x},t)$ standing for the
drift term and $D^{ij}({\bf x},t)$ the diffusion tensor, since we can take $%
P({\bf x},t)$ as a real positive and normalized function, such that it can
be interpreted as a (covariant) probability density. This probability
attribute of $P(x)$ is consistent with the fact that $P(x)$ can not be zero,
providing then that $g^{\mu \nu }(x)$ has an inverse, say $g_{\mu \nu }(x)$.

The Riemann space ${\cal {R}({G})}$ has been used to introduce the drift and
the diffusion tensor in Fokker-Planck dynamics. Regarding the metric $D^{ij}(%
{\bf x},t)$, the connection in the Euclidean part of ${\cal {R}({G})}$ is
given by 
\[
\Gamma _{\;k}^{ij}=\frac{1}{2}(\frac{\partial D^{jk}}{\partial x^{i}}+\frac{%
D^{ik}}{\partial x^{j}}-\frac{D^{ij}}{\partial x^{k}}).
\]
If $\Gamma _{k}^{ij}=0,$ we recover the Euclidean flat space, and therefore
there exists a transformation $U(x)$ such that 
\begin{equation}
UgU^{-1}=P(x)D\delta _{ij}dx^{j}{\otimes }dx^{i}-dx^{4}{\otimes }%
dx^{5}-dx^{5}{\otimes }dx^{4},  \label{galmetri2}
\end{equation}
where $D$ is a constant. Hence, the diffusion tensor can be diagonalized.
This is a result derived by Graham \cite{grah1} in a work analyzing the
invariance properties of the Fokker-Planck equation. In our case, the
invariance has been used, from the beginning, as a central ingredient to
write the Lagrangian given in Eq. (\ref{lagr1}) and the corresponding
covariant Eq. (\ref{f111}).

Let us briefly discuss the case of relativistic Fokker-Planck equation. This
can be obtained if we use Eq. (\ref{emb3}) with $x^{\nu }=(x^{i},\frac{ct}{%
\sqrt{2}},\frac{ct}{\sqrt{2}})$,where $c$ is the speed of light, and $f^{\nu
}(x)=(f^{i}({\bf x},t),\frac{P}{\sqrt{2}},\frac{P}{\sqrt{2}})$. In this
case, for instance, we have $x^{\mu }x_{\mu }=x^{i}x_{i}-(ct)^{2},$ which is
a vector in Minkowski space. Then we have the following correspondence of
5-tensors in ${\cal G}$ into 4-tensors in the Minkowski space,

\begin{eqnarray*}
\partial _{\mu } &\rightarrow &\partial _{\mu }=(\partial _{0}=\partial
_{ct},\partial _{i}), \\
S^{\mu } &\rightarrow &S^{\mu }=(S^{0}=P(x),\;S^{i}=D^{i}(x)P(x)+\partial
_{j}P(x)D^{ij}(x)), \\
\eta  &\rightarrow &\eta =\delta _{ij}dx^{j}{\otimes }dx^{i}-dx^{0}{\otimes }%
dx^{0}.
\end{eqnarray*}
Using these definitions and Eq. (\ref{f1}) (but now in the Minkowski space),
we derive a relativistic Fokker-Planck equation which has the same form as
Eq. (\ref{f21}) for $t\rightarrow ct$ (we consider $c=1$). Therefore, the
usual Fokker Planck equation can be taken as Lorentz invariant, provided
there exists a drift 4-vector given by $f^{\nu }(x)=(P({\bf x},t),f^{i}({\bf %
x},t))$, and a Riemannian metric given by 
\[
g=P(x)D_{ij}(x)dx^{j}{\otimes }dx^{i}-dx^{0}{\otimes }dx^{0}.
\]
In the next section we consider non-abelian symmetries.

\section{Non-Abelian Gauge Symmetry}

Generalization of these results for non abelian gauge fields can be
addressed as well. (From this point on, the covariant notation means
relativistic or non-relativistic theory.) In 5-dimensions a pure Yang-Mills
field can be described by the Lagrangian 
\begin{equation}
{\cal {L}}=-\frac{1}{4}F^{a\mu \nu }F_{a\mu \nu },  \label{lagr11}
\end{equation}
where the Latin index, $a$, stands for the gauge group, with generators $%
t^{a},\ \ a=1,2...n$, satisfying the Lie algebra $%
[t^{a},t^{b}]=C_{c}^{ab}t^{c},$ where $C_{c}^{ab}$ are structure constants
of the gauge group (sum over repeated Latin indices is assumed). The field
strength tensor $F_{\mu \nu }^{a}$ is given by 
\[
F_{\mu \nu a}=\partial _{\mu }J_{\nu a}-\partial _{\nu }J_{\mu a}-\lambda
C_{a}^{bc}J_{\mu b}J_{\nu c},
\]
for which the equation of motion is written as 
\[
D_{a}^{\mu b}F_{\mu \nu b}=0,
\]
where $D_{a}^{b\mu }$ is the covariant derivative given by $D_{a}^{\mu
b}=\partial ^{\mu }\delta _{a}^{b}+\lambda C_{a}^{bc}J_{c}^{\mu }.$ The
equations of motion for each component of $J$ are 
\begin{eqnarray}
\partial _{\nu }\partial _{\mu }J_{a}^{\mu } &=&\lambda C_{a}^{bc}\partial
_{\mu }(J_{c}^{\mu }J_{\nu b})+\lambda C_{a}^{bc}J_{c}^{\mu }\partial _{\mu
}J_{\nu b}  \nonumber \\
&+&\lambda C_{a}^{cb}J_{c}^{\mu }\partial _{\nu }J_{\mu b}+\lambda
^{2}C_{a}^{cb}C_{b}^{de}J_{c}^{\mu }J_{\mu d}J_{\nu e},  \label{fpym}
\end{eqnarray}
where use has been made of the aforementioned gauge condition $\partial
_{\mu }\partial ^{\mu }J_{\nu a}=0$. Despite the non-linear structure of
these equations, a Fokker-Planck system can be recognized, if we assume $J$
is defined as before (in terms of $S$), and discard all the non-linear terms
in Eq. (\ref{fpym}), such that $\partial ^{\mu }\partial _{\nu }J_{\mu a}=0$%
. As a consequence 
\begin{equation}
\partial _{\mu }J_{a}^{\mu }=\alpha ,  \label{fpym1}
\end{equation}
where $\alpha $ is a constant. Taking $\alpha =0$, we obtain Eq. (\ref{f1}),
and so a Fokker-Planck equation, for each gauge index $a$. On the other
hand, consider $\alpha <<1$, then Eq. (\ref{fpym}) reduces, up to second
order terms in $\lambda \alpha $, to 
\begin{equation}
\partial _{\nu }(\partial _{\mu }J_{a}^{\mu }+\lambda C_{a}^{bc}J_{c}^{\mu
}J_{\mu b})=2\lambda C_{a}^{bc}J_{c}^{\mu }\partial _{\mu }J_{\nu b}+\lambda
C_{a}^{bc}(\partial _{\nu }J_{c}^{\mu })J_{\mu b}.  \label{fpym2}
\end{equation}
The left-hand side of this equation can be integrated for each $\nu =1,...,5,
$ such the right side-hand results in a non-local term along each direction.
Discarding this non-local term we obtain the following nonlinear equation 
\begin{equation}
\partial _{\mu }J_{a}^{\mu }+\lambda C_{a}^{bc}J_{c}^{\mu }J_{\mu b}=0.
\label{fpym3}
\end{equation}

Let us consider as an example the $su(2)$ symmetry with $J_{a}^{\mu }$ \
defined by 
\begin{eqnarray*}
J_{a}^{i} &=&\epsilon _{aij}(D_{j}^{k}P_{k}+D_{j}^{nk}\partial _{k}P_{n})\ ,
\\
J_{a}^{4} &=&P_{a}\ , \\
J_{a}^{5} &=&0\ ,
\end{eqnarray*}
where both gauge and tensor indices are of the same tensor nature ($%
i,j,k,a,b,c=1,2,3$), $D_{j}^{k}=D_{j}^{k}(x)$ describes the drift term
(which is now a 2nd rank tensor, taking into account the vector and the
gauge index), whilst $D_{j}^{nk}$, independent of $x$, stands for the
diffusion term. Notice that this definition can be developed with the
reasoning used in the abelian case. With Eq. (\ref{fpym3}), we see that $%
\epsilon _{abc}J_{c}^{\mu }J_{\mu b}=\epsilon _{abc}J_{ic}J_{ib}=0.$ Hence 
\begin{equation}
\partial _{t}P_{a}=\epsilon _{aji}[\partial
_{i}(D_{j}^{b}P_{b})+D_{j}^{cb}\partial _{i}\partial _{b}P_{c}].  \label{f22}
\end{equation}

The content of this Fokker-Planck-like equation can be analyzed in a simple
particular situation. Consider the stationary situation $\partial
_{t}P_{a}=0,$ and define 
\begin{eqnarray*}
P_{2} &=&P_{3}=P, \\
D_{2}^{1} &=&kx^{3}=kz, \\
D_{3}^{1} &=&kx^{2}=ky, \\
D_{2}^{13} &=&D_{3}^{12}=\frac{D}{2},
\end{eqnarray*}
where $P,D$ and $k$ are constant and the other components of $D_{j}^{b}$ and 
$D_{j}^{cb}$ are zero. (The expressions for $D_{2}^{1}$ and $D_{3}^{1},$ the
drift terms, and $D_{2}^{13}=D_{3}^{12},$ the diffusion tensor assure we
have an Ornstein-Uhlenbeck-like process \cite{risken,gard1} for this color
theory.) Writing $P_{1}(y,z)=\varphi (y)\phi (z),$ we derive 
\[
\frac{1}{\phi (z)}\left[ \frac{D}{2}\frac{d^{2}}{dz^{2}}\phi (z)+\frac{d}{dz}%
(kz\phi (z))\right] =\frac{1}{\varphi (y)}\left[ \frac{D}{2}\frac{d^{2}}{%
dy^{2}}\varphi (y)+\frac{d}{dy}[kz\varphi (y)]\right] .
\]
Therefore, we can write 
\[
\frac{D}{2}\frac{d^{2}}{dy^{2}}\varphi (y)+\frac{d}{dy}[ky\varphi
(y)]=\varphi (y)F,
\]
where $F$ is a constant. If $F\neq 0,$ a solution is given with $F=3k,$ such
that 
\[
\varphi (y)=a_{0}(y^{2}+\frac{D}{k})+a_{1}\exp (-ky^{2}/D)\ ,
\]
where $a_{0},a_{1}\ $\ are constant; and similarly for $\phi (z)$. However,
these type of solution diverges for $y,z\rightarrow \infty .$ A
non-divergent solution is found for $F=0.$ In this case we recover a known
result for $P_{1}(y,z)$, that is 
\begin{eqnarray*}
P_{1}(y,z) &=&\varphi (y)\phi (z) \\
&=&\frac{1}{N}\exp (-k(y^{2}+z^{2})/D),
\end{eqnarray*}
with $N$ being a normalization constant.

\section{Concluding Remarks}

Summarizing, our main goal in this work has been to derive the Fokker-Planck
dynamics via a variational principle, considering gauge invariant
Lagrangians written in a (4+1) Minkowski space ${\cal G}.$ First we consider
the $U(1)$ symmetry with a suitable choice for the gauge condition, and a
Riemannian manifold ${\cal R(G)}$ specified by the metric $g$ given in Eq. (%
\ref{metric11}). Then using Eq. (\ref{f1}) the Fokker-Planck equation has
been derived. The physical meaning of the manifold ${\cal {R}({G})}$ is
crucial for the definition of the diffusion and drift terms, which are
derived from the metric tensor $g$. This is why the equivalence principle
was no longer invoked.

The analysis has been extended to deal with non-abelian symmetries and the
relativistic case, taking the advantage of proper embeddings in this
5-dimensional formalism. For non-abelian groups, despite the difficulties
imposed with the non-linearity, we have been able to recognize a
Fokker-Planck dynamics and discuss, as an example, a solution for the $su(2)$
gauge symmetry. Other possibilities for the gauge group remains to be
studied, and  a situation involving the $su(3)$ symmetry will be addressed
elsewhere.

\noindent {\bf Acknowledgments} We thank M. Revzen and A. Zelnikov for
stimulants discussions, and NSERC (Canada) for partial financial support.

\end{document}